\documentclass[12pt,preprint]{aastex}
\newcommand{\nraoblurb}{The National Radio Astronomy Observatory is
a facility of the National Science Foundation operated under cooperative
agreement by Associated Universities, Inc.}
\newcommand{\te}{\ensuremath{{T_{e}}}}
\newcommand{\tc}{\ensuremath{{T_{C}}}}
\newcommand{\lte}{\ensuremath{{T_{e}^{\ast}}}}
\newcommand{\rmxaax}{RMxAA}

\newcommand{\kpc}{\ensuremath{\,{\rm kpc}}}

\newcommand{\ghz}{\ensuremath{\,{\rm GHz}}}
\newcommand{\kel}{\ensuremath{\,{\rm K}}}
\newcommand{\percc}{\ensuremath{\,{\rm cm^{-3}}}}

\newcommand{\kms}{\ensuremath{\,{\rm km\,s}^{-1}}}
\newcommand{\rgal}{\ensuremath{{R_{\rm gal}}}}
\newcommand{\dsun}{\ensuremath{{d_{\rm sun}}}}
\newcommand{\ro}{\ensuremath{{R_{\rm 0}}}}

\newcommand{\kkpc}{${\,{\rm K\, kpc^{-1}}}$}

\newcommand{\hi}{{\rm H\,}{{\sc i}}}
\newcommand{\hii}{{\rm H\,}{{\sc ii}}}       
\newcommand{\he}[1]{$^{#1}{\rm He}$}

\newcommand{\cii}{{\rm C\,}{{\sc ii}}}
\newcommand{\oiii}{{\rm O\,}{{\sc iii}}}
\newcommand{\oii}{{\rm O\,}{{\sc ii}}}

\newcommand{\nii}{{\rm N\,}{{\sc ii}}}
\newcommand{\siii}{{\rm S\,}{{\sc iii}}}
\newcommand{\ariii}{{\rm Ar\,}{{\sc iii}}}

\newcommand{\hefour}{\ensuremath{{}^4{\rm He}}}

\newcommand{\expo}[1]{\ensuremath{10^{#1}}}

\slugcomment{Accepted for publication in the ApJ}

\shorttitle{The Galactic Electron Temperature Gradient}
\shortauthors{Quireza et al.}

\begin{document}

\title{The Electron Temperature Gradient in the Galactic Disk}

\author{Cintia Quireza\altaffilmark{1,2,3}, Robert T. Rood\altaffilmark{3}, 
T. M. Bania\altaffilmark{4}, \\ Dana S. Balser\altaffilmark{5} \&
Walter J. Maciel\altaffilmark{2}}

\altaffiltext{1}{Observat\'orio Nacional, Rua General Jos\'e Cristino 77, 
20921-400, Rio de Janeiro, RJ, Brazil (quireza@on.br).}
\altaffiltext{2}{Instituto de Astronomia, Geof\'{\i}sica e Ci\^encias 
Atmosf\'ericas (IAG), Universidade de S\~ao Paulo, Rua do Mat\~ao 1226, 
05508-900, S\~ao Paulo, SP, Brazil.}
\altaffiltext{3}{Astronomy Department, University of Virginia, P.O.Box 3818, 
Charlottesville VA 22903-0818, USA.}
\altaffiltext{4}{Institute for Astrophysical Research, Department of Astronomy,
Boston University, 725 Commonwealth Avenue, Boston MA 02215, USA.}
\altaffiltext{5}{National Radio Astronomy Observatory, P.O. Box 2, 
Green Bank WV 24944, USA.}

\begin{abstract}

We derive the electron temperature gradient in the Galactic disk using
a sample of \hii\ regions that spans Galactocentric distances 0--17
kpc. The electron temperature was calculated using high precision
radio recombination line and continuum observations for more than 100
\hii\ regions. Nebular Galactocentric distances were calculated in a
consistent manner using the radial velocities measured by our radio
recombination line survey. The large number of nebulae widely
distributed over the Galactic disk together with the uniformity of our
data provide a secure estimate of the present electron temperature
gradient in the Milky Way. Because metals are the main coolants in the
photoionized gas, the electron temperature along the Galactic disk
should be directly related to the distribution of heavy elements in
the Milky Way. Our best estimate of the electron temperature gradient
is derived from a sample of 76 sources for which we have the highest
quality data.  The present gradient in electron temperature has a
minimum at the Galactic Center and rises at a rate of 287 $\pm$ 46
\kkpc.  There are no significant variations in the value of the
gradient as a function of Galactocentric radius or azimuth.  The
scatter we find in the \hii\ region electron temperatures at a given
Galactocentric radius is not due to observational error, but rather to
intrinsic fluctuations in these temperatures which are almost
certainly due to fluctuations in the nebular heavy element abundances.
Comparing the \hii\ region gradient with the much steeper gradient
found for planetary nebulae suggests that the electron temperature
gradient evolves with time, becoming flatter as a consequence of the
chemical evolution of the Milky Way's disk.

\end{abstract}


\section{INTRODUCTION}

Churchwell \& Walmsley (1975) pioneered \hii\ region radio
recombination line (RRL) studies of the relationship between nebular
electron temperatures, \te, and Galactocentric distance, \rgal,
(Churchwell et al. 1978; Wink et al. 1983; Shaver et al. 1983; and
others).  Because RRLs are not obscured by interstellar dust,
relatively faint \hii\ regions at extremely large distances from the
Sun could be detected.  They found that there was a Galactic
temperature gradient wherein \te\ is low in the Galactic Center and
increases with \rgal.  Such a gradient was first observed in other
nearby spiral galaxies by Searle (1971), Rubin et al. (1972) and Smith
(1975).  This Milky Way electron temperature gradient was confirmed by
radio continuum emission (Omar et al. 2002) and by [\oiii]\ forbidden
line optical observations (Peimbert et al. 1978; Deharveng et
al. 2000).  Because heavy elements cool photoionized gas, \hii\ region
electron temperatures are directly related to the heavy element
abundance: low \te\ corresponds to higher heavy element abundances
because of the greater cooling rate and vice versa.  Consequently,
\te\ gradients should be inversely related to metal abundance
gradients (but see \S 6).  Because there has been more stellar
processing in the inner Galaxy one expects that on average the
metallicity decreases as a function of \rgal.

The existence of a Galactic gradient in \hii\ region electron
temperature is now firmly established. Nevertheless, there are still
uncertainties in the magnitude of the gradient and the possible
existence of real variations, both in \rgal\ and Galactic azimuth, of
nebular \te.  Extant RRL studies yield \te\ gradients that, roughly,
vary from 250--440 \kkpc.  Discrepancies in the results obtained from
different studies may be attributed to several factors, including the
source sample, the \te\ derivation, etc.  Since the exact value
of the \te\ gradient provides an important constraint on models for
Galactic chemical evolution, it must be determined as accurately as
possible.

Here we derive \hii\ region electron temperatures based on radio
recombination line and continuum data for a large sample of nebulae
widely distributed across the Galactic disk. The RRL data are of
unprecedented sensitivity compared with previous studies.  We examine
anew the Galactic temperature gradient and assess the magnitude and
origin of the \te\ dispersion at a given \rgal. In \S\,\ref{sec:obs}
we describe our \hii\ region sample. In \S\,\ref{sec:te} we derive the
nebular electron temperatures and discuss non-LTE effects.  Nebular
Galactocentric distances, \rgal, and heliocentric distances, \dsun,
are derived in \S\,\ref{sec:dist}.  In \S\,\ref{sec:tegrad} our new
determination of the electron temperature gradient in the Galactic
disk is made.  Here we also investigate a possible spatial variation
of this gradient.  We discuss the astrophysical implications of our
efforts in \S\,\ref{sec:discuss}.

\section{OBSERVATIONAL SAMPLE} \label{sec:obs}

Our \hii\ region sample data are described by Quireza et al. (2006a)
and references therein.  The radio recombination line and continuum
data result from two different experiments.  Neither was targeted to
investigate radial gradients in the Galactic disk.  The first one
(hereafter $^3$He survey) is a study of the abundance of $^3$He in the
Milky Way interstellar medium (Rood et al. 1984; Bania et al. 1987,
1997; Balser et al. 1994) using the hyperfine transition of $^3$He$^+$
at 8.665 GHz.  The observing techniques required that a number of
radio recombination lines be measured simultaneously with the
$^3$He$^+$ transition. These RRLs were used both to monitor the system
performance and also to measure the spectral baseline frequency
structure.  As a consequence of the large integration times (over 100
hours in many cases) accumulated during this experiment, we obtained
extremely high sensitivity measurements of the He\,91\,$\alpha$ and
H\,91\,$\alpha$ ($\Delta n = 1$) RRL transitons for a significant
number of Galactic \hii\ regions.  The second experiment (hereafter
\cii\ survey) is a study of \cii\ recombination lines in
photo-dissociation (PDR) regions surrounding the \hii\ regions.  For
this survey we simultaneously observed the H, He, and C 91\,$\alpha$
and 92\,$\alpha$ RRL transitions.  Since recombination lines of the
same order with similar principal quantum numbers, such as
H\,91\,$\alpha$ and H\,92\,$\alpha$ or He\,91\,$\alpha$ and
He\,92\,$\alpha$, should have the same intensity, we averaged the
91\,$\alpha$ and 92\,$\alpha$ spectra to attain higher sensitivity
(see Quireza et al. 2006a for details).
 
All observations were made near 8.6 GHz (3.5 cm) with the National
Radio Astronomy Observatory\footnote{\nraoblurb } (NRAO) 140 Foot (43
m) telescope in Green Bank, WV, which has a half power beam width
(HPBW) of $3\farcm20$ at this frequency. Our sample has 106 sources:
47 nebulae from the $^3$He survey and 66 from the \cii\ survey.
(There are 7 objects in common.)

The quantities needed to derive nebular electron temperatures are the
H\,91\,$\alpha$ line peak intensities, $T_L$, the line full width at
half-maximum, $\Delta v$, and the continuum intensity, $T_C$. These
parameters and their errors may be found in Quireza et al. (2006a).
There we also assess the quality of our data: we have defined quality
factors, QFs, for both spectral line and continuum data. Because we
cannot quantify the systematic errors, these QFs provide a qualitative
measure of them.  By choosing subsets of data with different QF's we
can assess the importance of systematic errors on our conclusions
reported here. Quality factor {\it A} sources refer to our best data:
the spectra are almost noiseless and systematic errors appear to be
neglibible.  The QF decreases from {\it A} to {\it E}.  We judge QFs
{\it D} and {\it E} to be of too low confidence to be included in our
analysis here.

\section{DETERMINATION OF THE PHYSICAL PARAMETERS} \label{sec:te}

\subsection{LTE Electron Temperature} \label{sec:lte}

The electron temperature of each nebula is derived from the observed
radio recombination line-to-continuum ratio, $T_L/T_C$.  We model our
sources as homogeneous, isothermal spheres.  This approximation allows
us to calculate the electron temperature without having to know the
distance to the source.  If the distance is known many other nebular
physical properties, such as electron density and emission measure,
can be derived.

Most of the emission observed from \hii\ regions is continuum
radiation produced by free-free thermal Bremsstrahlung in the
plasma. At high frequencies the nebular gas is optically thin and the
ratio between the brightness temperature of a recombination line and
that of the free-free emission continuum depends of the radio
frequency and the gas temperature, but is independent of the electron
density, $n_e$.  Thus the observed $T_L/T_C$ may be used to estimate
the electron temperature of the \hii\ region (Goldberg 1968; Rohlfs \&
Wilson 2000).  Assuming local thermodynamic equilibrium (LTE) and
negligible pressure broadening of the lines by electron impacts, the
electron temperature is given by:
\begin{equation}\label{eq:lte}
{\left(T_e^{\ast} \over K\right)} = \left[7103.3 {\left(\nu_L
\over {\rm GHz}\right)}^{1.1} 
\left({T_C \over\strut T_L ({\rm H{^+}})} \right)
{\left(\Delta v ({\rm H{^+}}) \over \kms \right)}^{-1} 
\left( 1 + {\strut n({^4}{\rm He}^+)\over\strut n({\rm H}^+)}
\right)^{-1} \right]^{0.87}
\end{equation}
\noindent where we distinguish the LTE temperature, $T_e^{\ast}$, from
the electron temperature corrected for non-LTE and high-density
effects, $T_e$. The line frequency, $\nu_L = 8.584823$ GHz,
corresponds to the rest frequency of the H\,91\,$\alpha$ recombination
line.  Here $T_C$\,(K) is the continuum antenna temperature; $T_L$\,(K) and
$\Delta v$\,(\kms) are the H\,91\,$\alpha$ recombination line antenna 
temperature and FWHM linewidth.  
The $n({^4}{\rm He}{^+})/n({\rm H}{^+})$ ionic abundance ratio was
calculated using the areas of Gaussian fits to the H and He
recombination lines (Peimbert et al. 1992):
\begin{equation}\label{eq:hephp}
{n({^4}{\rm He}{^+}) \over n({\rm H}{^+})} = {{T_L ({\rm
{^4}He{^+}})\, \Delta v ({\rm {^4}He{^+}})}\over{T_L ({\rm H{^+}})\,
\Delta v ({\rm H{^+}})}}.
\end{equation}
\noindent For a small number of objects we do not have good
measurements of the \hefour\ transition.  In these cases we used a
constant value of $0.07 \pm 0.02$ for the $n({^4}{\rm He}{^+})/n({\rm
H}{^+})$ abundance ratio.  This ratio is the average of our 80 best QF
({\it A}, {\it B}, and {\it C}\/) sources and is typical for \hii\
regions (Churchwell et al. 1974; Shaver et al. 1983).

The nebular LTE electron temperatures and their errors, $\sigma_\te$,
together with the Galactocentric distances, \rgal, heliocentric
distances, \dsun, and the $n({^4}{\rm He}{^+})/n({\rm H}{^+})$ ionic
abundance ratios, are listed in Table~\ref{tab:physparam}. Also given
are the name of the source and some of its physical properties
including the spherical angular size, $\Theta_{\rm diam}$,  
linear diameter, $D$, 
flux density, $S_\nu$,
continuum brightness temperature, ${T_{\rm C}^{\rm B}}$, 
and electron density, $n_{\rm e}$, 
(see the discussion below).  Each nebula's survey membership 
(either $^3$He or \cii) is indicated.

The electron temperature errors were derived by propagating the
Gaussian fitting errors for the line and continuum measurements. These
$\sigma_\te$ errors vary from 0.3--17.8\% (2.2\% on average).  For our
best QF data, $\sigma_\te$ errors vary from 0.4--11.0\% (1.3\% on
average). These are lower limits to the temperature errors.  Due to
baseline problems and complex nebular structures, uncertainties in
continuum measurements are certainly larger than the one we estimate,
reaching 10\% or even 20\% in the worst cases. Since we have no way to
quantify systematic uncertainties, we use the quality factors to
estimate the effect of the systematic errors on the electron
temperatures. The QFs for continuum and spectral line parameters are
listed, in this order, in Table~\ref{tab:physparam}.

The continuum observing mode can also affect the uncertainty in the
measurements.  We made continuum measurements using both the switched
power (SP) and total power (TP) techniques (see Quireza et al. 2006a).
There we point out that as a source was tracked across the sky we
interleaved the recombination line and SP continuum measurements such
that both the continuum and line data span the same hour angle ranges
and experience the same weather conditions.  Thus the line and
continuum data would need identical corrections for telescope gain and
atmospheric opacity.  When calculating source properties that depend
on the line-to-continuum ratio, such as the electron temperature,
Quireza et al. suggest using SP continuum measurements.  We therefore
normally use SP continuum observations to calculate \lte\ because many
telescope effects are canceled in the line-to-continuum ratio.  We did
not have SP continuum measurements for a small number of our sources
and had to use TP continuum data to calculate \lte.  We estimate that
the use of TP continuum data should give at most a $\sim\,$20\%
uncertainty in \lte.  The observing mode used is listed in
Table~\ref{tab:physparam} for each nebula.

Accurate calculations of $T_{e}$ should incorporate non-LTE
effects including departures from LTE, stimulated emission, and
pressure broadening from electron impacts.  These corrections for
non-LTE effects are sensitive to the local density and thus the \hii\
region geometry.  Nevertheless, it has been shown that under many
conditions LTE is a good approximation and the LTE electron
temperature $T_{e}^{*}$ is close to $T_{e}$.

For conditions close to LTE and where pressure broadening is not
significant, the $\beta/\alpha$ RRL intensity ratio should be about
0.28, and the $\beta/\alpha$ RRL line width ratio should be close to 1
(Shaver \& Wilson 1979).  We determined these ratios for the $^3$He
survey where the H\,114\,$\beta$ line has been observed.  For our best
QF data (28 objects), the average intensity ratio is $0.26 \pm 0.03$
and the average line width ratio is $1.02 \pm 0.06$.  Moreover, Shaver
(1980) defined a RRL observing frequency, $\nu_{\rm LTE}$, such that
\lte\,=\,\te. This optimal radio recombination line frequency is a
function of emission measure: $\nu_{\rm LTE} \sim 0.081 EM^{0.36}$.
It is essentially independent of density, temperature, or structure
within the nebula.  Using our best QF data (88 objects) we obtain an
average frequency $\nu_{\rm LTE} = 6.9 \pm\ 3.6$\ghz.  This is close
to our observing frequency of 8.584\ghz.  Furthermore, detailed
density structure models for a subset of our $^3$He survey nebulae
were made using high resolution VLA radio continuum images and
high-order RRLs that were sensitive to local electron densities
(Balser et al. 1999). The non-LTE electron temperatures calculated for
these models are very close to the LTE electron temperature determined
using the H91$\alpha$ lines.  We therefore conclude that our LTE
electron temperature must be very close to the real average nebular
electron temperature.

\subsection{Nebular Angular Size} \label{sec:size}

We derived the \hii\ region angular sizes by assuming homogeneous,
spherical nebulae.  Each nebula has an observed full width at
half-maximum size of $\Theta_{\rm src} = {[\Theta(RA)\,
\Theta(DEC)]}^{1/2}$ (geometric mean, see Quireza et al. 2006a).
Assuming that the source has a Gaussian brightness distribution and
that the telescope beam has a Gaussian pattern with half power beam
width, HPBW, then $\Theta_{\rm src}$ is:
\begin{equation}\label{eq:size}
\Theta_G = \root \of {({\Theta_{\rm src}}^2 - HPBW^2)}\ ,
\end{equation}
\noindent where $\Theta_G$ is the nebular Gaussian angular size
(Mezger \& Henderson 1967).  We use the technique developed by Panagia
\& Walmsley (1978) to derive $\Theta_{\rm diam}$, the angular size of
a homogeneous, spherical nebula from $\Theta_G$.  We use TP continuum
measurements if at all possible.  The nebular $\Theta_{\rm diam}$ so
derived is listed in Table~\ref{tab:physparam} together with the
linear diameter, $D$.  Objects whose angular sizes are close to the
telescope HPBW are not well resolved and consequently have less
precise estimates for their sizes.  This will give a larger error for
any physical property whose derivation depends on the angular size.  
One important property which requires precise angular
size measurements is the electron density.  


\subsection{R.M.S. Electron Density} \label{sec:ne}

If a homogeneous, spherical nebula is also optically thin, then the
electron density can be calculated from the peak continuum brightness
temperature of the source, ${T_{\rm C}^{\rm B}}$ (Balser 1995):

\begin{equation}\label{eq:ne}
{n_e}\,(\percc\,) = \left[ {{{T_{\rm C}}^{\rm B}\,({\rm K})\, 
{\nu}\,({\rm GHz})^2\, \lte\,({\rm K})^{1/2}} 
\over {8.77 \times 10^{-3}\, \ln (X)\, 
\Theta_{\rm diam}\,({\rm arcmin})\,\dsun\,(\kpc\,)\,}} \right]^{1/2}\, ,
\end{equation}
\noindent where 
$\nu$ is the frequency (8.66565 GHz),
\lte\ is the electron temperature,
$X = 4.954 \times 10^{-2}\, {{T_e^{\ast}\,}^{3/2} / \nu}$,
$\Theta_{\rm diam}$ is the spherical size, and 
\dsun\ is the source's distance from the Sun (see \S\,6).
Here we assume no doubly ionized helium gas within the \hii\ region.
The flux density, $S_\nu$, is:
\begin{equation}\label{eq:flux}
S_\nu\, (Jy) = 1.223\, T_{\rm C}^{\rm B}\, (K) {\left(
\Theta_{\rm diam} \over \lambda \right)}^2 =
2.647 \left({T_{\rm C} \over \eta_{\rm b}}\right) {\left(
\Theta_{\rm src} \over \lambda \right)}^2,
\end{equation}
where temperatures are in K, angular sizes are in arcmin, and the
wavelength is in cm.  The beam efficiency of the 140 Foot telescope is
$\eta_{\rm b} = 0.86$.  

The nebular flux and electron densities calculated in this way are
listed in Table~\ref{tab:physparam}.  Because $T_C^B$ measures the
continuum radiation produced by free-free thermal Bremsstralung in the
plasma its value depends on the integral of $n_e^2$ along the line of
sight. Thus the electron densities derived here are the root mean
square (rms) density of the entire \hii\ region. If the nebula is
homogeneous, then the rms electron density equals the local
density. Detailed analysis, however, shows that this is not generally
true for Galactic \hii\ regions (Osterbrock \& Flather 1959).  The rms
densities are probably somewhat lower than the true densities of the
emitting regions, since the gas is not evenly distributed throughout
the \hii\ region (Fich \& Silkey 1991).

\subsection{Comparison with Previous Studies} \label{sec:compte}

Here we compare our electron temperatures with results from the
literature for nebulae in our sample.  The majority of these previous
efforts were also studying the Milky Way \te\ gradient.  Most \te\
estimates come from RRLs, although measurements in other spectral
regions are available for a limited number of objects.  We summarize
the \te\ differences in Table~\ref{tab:tecomp2} which lists the number
of sources in common, $N$, the average percentage difference between
our results and the reference, and the average percentage of the
absolute value of difference between our results and the reference.
The average difference will reveal any offset in \te\ scales; the
average absolute value difference measures the scatter between the
studies.  The table also lists the type of observation and the method
used to calculate \te\ (line-to-continuum ratio; optical forbidden
lines; radio continuum emission; etc).

\begin{figure}
\epsscale{0.75} \plotone{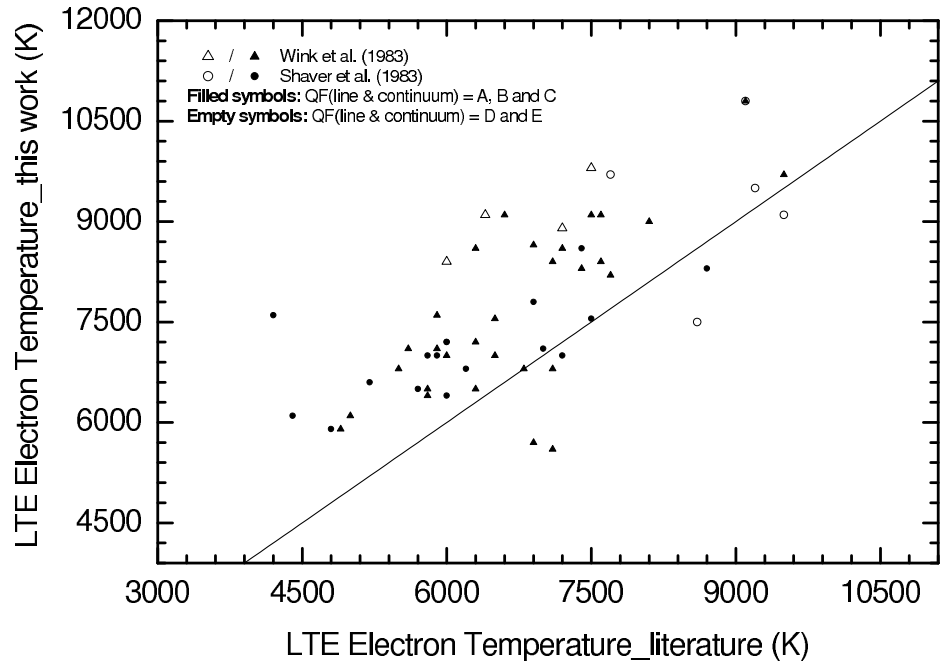} \figcaption[f3.comp.eps]
{Nebular LTE electron temperatures derived here compared with those 
from Shaver et al. (1983) (circles) and Wink et al. (1983) (triangles)
for objects in common. Filled symbols denote our highest quality data 
(QFs {\it A}, {\it B}, \& {\it C}\/); open symbols flag our poorer 
quality data.  The locus of \te(us) = \te(literature) is shown by
the solid line.
\label{fig:comp2}}
\end{figure}

It is immediately apparent that our \te's are systematically high
compared with previous studies. While one might expect systematic
differences between \te's determined by different methods, the largest
offsets are found between our values and those of Wink et al. (1983)
(WBW83) and Shaver et al. (1983) (SMNDP83) which are both RRL studies
similar to ours with many sources in common. In Figure~\ref{fig:comp2}
we compare our LTE electron temperatures with their \te\,'s.  Our best
QF data ({\it A}, {\it B}, and {\it C}\/) are plotted as filled
symbols. Our values are systematically 11\% higher than WBW83 and 13\%
higher than SMNDP83.

We considered the origin of these differences in some detail.  Each
and every term in Eq.~\ref{eq:lte} may be the explanation.  There may
also be issues of calibration between line and continuum measurements
and between telescopes.  All the RRL studies were made at different
frequencies using different telescopes with correspondingly different
beam sizes.  Thus each survey probes different volumes of each nebula.
Certainly there is complex structure inside some \hii\ regions which
includes large density and temperature fluctuations.  The \te\ derived
from disparate RRL transitions using different telescopes can well be
different, especially under the assumption of LTE.

The beam size for the WBW83 sudy was substantially smaller than ours,
so on the average they were observing higher density gas than we
did. Exactly how to interpret this is quite complex and would require
detailed density and excitation modelling for each nebula.  The
continuum measurement technique is also important.  These details can
lead to \tc\ measurements that are too low (the continuum data do not
extend beyond the \hii\ region) or too high (Galactic non-thermal
continuum is included in \tc).

In Balser et al. (1999) our continuum observations from the 140 Foot,
MPIfR 100 m, and the VLA were carefully cross-calibrated, confirming
the calibration techniques used for the \he3\ and \cii\ surveys.
Furthermore, our recombination line data are vastly superior to these
previous studies due to a combination of improved technology (system
temperatures of $\sim$\,35\,K compared with $\sim$\,100\,K) and longer
integration times (tens of hours compared with tens of minutes).  In
sum, we have yet to identify just why WBW83 and SMNDP83 have lower
\te\ values.  

%
%

We have in fact reason to believe that our electron temperatures
are the best values derived to date.  We have substantially better
recombination line data that have unsurpassed sensitivity due to a
combination of modern receivers and our extremely long integration
times.  The modes of the distributions of source integration times for
the carbon and \he3\ surveys are $\sim$\,15 hrs and $\sim$\,50 hrs,
respectively.  Previous efforts have source integration times of
$\sim$\,30 min or less.  Finally, these extremely high signal to noise
spectra allow us to model the spectral baseline frequency structure
with unprecidented accuracy which gives us great confidence in our
determination of the recombination line parameters.

\section{NEBULAR DISTANCES} \label{sec:dist}

We derived the Galactocentric distance, \rgal, for each nebula using
the observed recombination line LSR velocity and assuming a Galactic
rotation curve.  For sources located inside the solar orbit we used
the Clemens (1985) rotation curve; otherwise the Brand \& Blitz (1993)
rotation curve was used.  Both rotation curves assume purely circular
rotation laws and they place the Sun at a Galactocentric distance of
$R_0 = 8.5$ kpc orbiting the Galactic Center at an LSR circular
velocity of $\Theta_0 = 220$ km s$^{-1}$.  The nebular RRL velocities
can be found in Table~2 of Quireza et al. (2006a).  Here we have
thus chosen to use only these kinematic determinations of the \rgal\ 
of our sources.  Many of the previous studies of the Milky Way 
\te\ gradient use a mix of techniques to establish the nebular \rgal.

The nebular heliocentric distance, \dsun, is also listed in Table
\ref{tab:physparam}.  These distances were also derived kinematically
from the observed nebular LSR velocity.  For sources located inside
the solar orbit, each radial velocity value corresponds to two
distances ( the ``near'' and ``far'' kinematic distances) equally
spaced on either side of the tangent point. In most cases, we were
able to resolve the kinematic distance ambiguity by measuring the 21
cm \hi\ absorption spectrum toward the nebular continuum.
Discrimination between the near and far distance was done by comparing
the maximum velocity of the \hi\ absorption with that of the \hii\
region recombination line.  For those objects for which we could not
resolve the ambiguity via \hi\ absorption, we used distances available
in the literature.  Our \hi\ survey, observational technique, method
of analysis, and detailed description of our \dsun\ derivations are
described by Quireza et al. (2006b).

\begin{figure}
\epsscale{0.75} \plotone{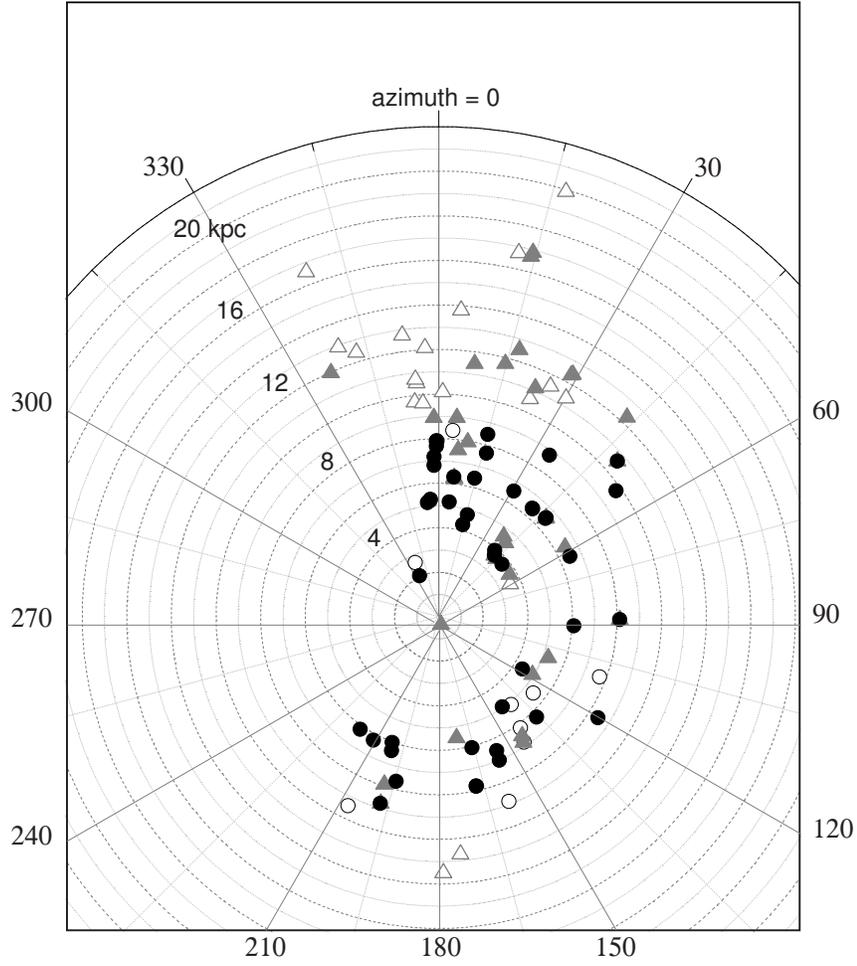} \figcaption[f4.azm.eps]
{Distribution of the \hii\ region sample projected onto the Galactic 
plane plotted as a function of Galactocentric radius and azimuth. 
Different symbols identify the $^3$He (triangles) and \cii\ survey 
(circles) nebulae.  Filled/open symbols have the same meaning as in 
Figure~\ref{fig:comp2}.
\label{fig:azm}}
\end{figure}

Figure~\ref{fig:azm} shows the distribution of our \hii\ region sample
projected onto the Galactic plane. Only sources with known \dsun\ are
plotted. The majority of our \hii\ regions are located in the first
and fourth Galactic quadrants and their \rgal's can reach $\sim $19
kpc. Some sources are located beyond the Galactic Center with
\dsun\,$\sim$\,20 kpc. Different symbols identify the $^3$He
(triangles) and \cii\ survey (circles) nebulae.  The $^3$He sources
span a larger range of Galactocentric radius and azimuth than do the
\cii\ nebulae.  Filled symbols denote our highest QF sources (QFs {\it
A}, {\it B}, \& {\it C}\/); open symbols flag our poorer quality data.
Unfortunately, many of the low QF objects are astrophysically
significant; they are located in areas of the disk not well covered by
nebulae with more accurate data.

%
%
There is good reason for using only kinematic distances:
homogeneity of approach.  A detailed comparison between optical and
kinematic distances is beyond the scope of this paper.  Both approaches
have their limitations and perhaps the resolution will be a direct
trignometic measurement using VLBI techniques which is being pioneered
by Mark Reid and his team (Xu et al. 2006; Hachisuka et al. 2006).

It is common to cite optical spectrophotometric distance
uncertainties to be of order 15\%.  But as one delves into the
literature in detail one finds a greater dispersion.  For any given
analysis the 15\% holds, but slight differences in spectral
classification, adopted luminosity scale which is model atmosphere
dependent, and uncertainties in the extinction correction conspire to
give a larger uncertainty when different research groups measure the
distances to the same object.  These systematic errors are the optical
counterpart of the streaming motion uncertainty inherent in kinematic
distances.  So the consistency of optical distances may not be all that
much better than the dispersion between optical and radio distances.
Most of our \hii\ regions at not optically visible.  A quick web
search for the 7 most distant (in \rgal) nebulae shows that the mean
absolute value of the radio-optical discrepancy for these sources is
11\% $\pm$ 6\%.

Non-circular streaming motions will certainly affect the
kinematically derived distances.  If we knew what the Galactic scale
streaming motions are we would certainly correct for them.  The Galactic
Bar, the spiral density wave, accretion events such as the Sgr dwarf
elliptical galaxy, and close encounters such as that which occured for
the LMC will all generate large-scale, asymmetric streaming motions in
the Milky Way disk.  At present there is no way to calculate these
effects accurately and thus generate the true velocity field of the
Galaxy.  

\section{ELECTRON TEMPERATURE GRADIENT} \label{sec:tegrad}

\subsection{Gradient Magnitude} \label{sec:magtegrad}

Our \hii\ region sample contains some sources observed in both the
$^3$He and \cii\ surveys as well as nebulae that have RRL spectra
taken toward multiple positions.  For sources common to both
surveys\footnote{G16.936$+$0.75 (M16), G23.421$-$0.21, G25.382$-$0.17
(3C385), G43.169$+$0.0 (W49), G49.384$-$0.29, G49.486$-$0.38 (W51)
and G79.293$+$1.29}, the \lte\ and \rgal\ values derived for the
$^3$He and \cii\ surveys were averaged (simple mean).  These nebulae
do {\it not} appear twice in the analysis of the electron temperature
gradient.  We only averaged data of good QFs ({\it C} or better) and
which also were in good agreement.  (The origin of some differences
between the two surveys is discussed by Quireza et al. [2006a].)
These \lte\ values are consistent with each other within the errors;
differences are not larger than $\sim$\,10\%.

In the $^3$He survey five morphologically complex \hii\ regions had
spectra taken toward two or three different positions. These nebulae
are:
G16.936+0.75 (the brightest and northern most components:
M16 and M16\,N), 
M17 (northern and southern most components: M17\,N and M17\,S),
NGC\,6334 (NGC\,6334\,A and NGC\,6334\,D), 
Rosette (Rosette\,A and Rosette\,B) and 
S209 (brightest, northern and southern most components: S209, 
S209\,N and S209\,S). 
Because these nebulae are extended objects with angular diameters
larger than the 3\farcm20 HPBW beam size, each of the observed
components is included in our analysis of the \lte\ gradient.  Since
these components are separated by more than a beamwidth any \lte\
differences within an object are real temperature fluctuations.

\begin{figure}
\epsscale{0.75} \plotone{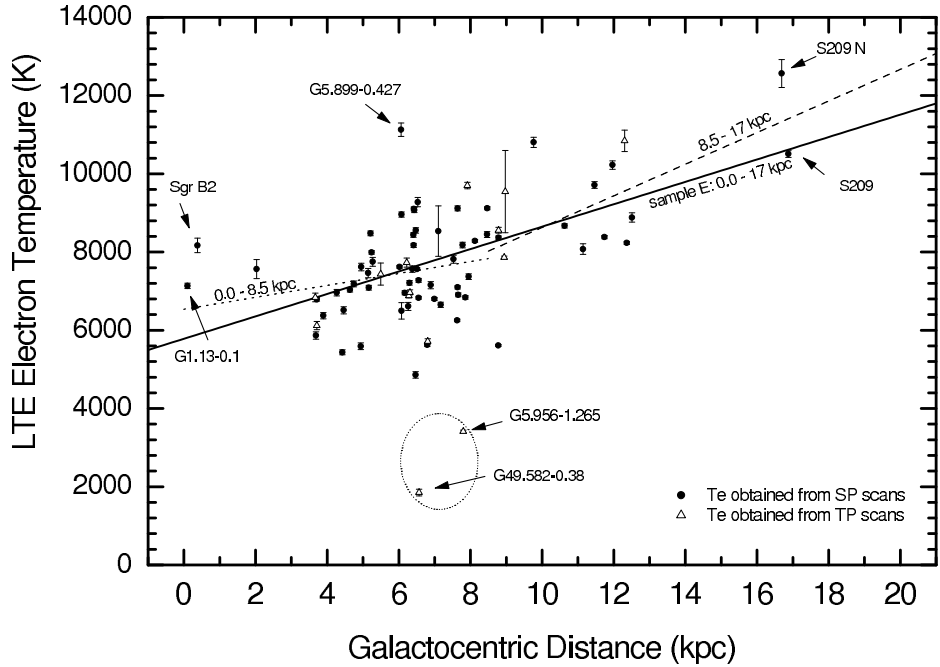} \figcaption[f5.tegrad.eps]
{Nebular LTE electron temperature plotted as a function of the 
Galactocentric distance for the best data (QFs {\it C} or better --- 
Sample B of Table~\ref{tab:tefit}).
Least squares linear fits to the gradient for the entire sample and, 
separately, for nebulae located inside and outside the solar orbit 
are shown.  The gradient is flatter for \hii\ regions in the 
inner Milky Way.
\label{fig:tegrad}}
\end{figure}

Figure~\ref{fig:tegrad} shows our nebular LTE electron temperatures,
\lte, plotted as a function of the Galactocentric distance, \rgal.  We
include only the 78 sources with our best data (QFs = {\it A}, {\it B}, 
\& {\it C} --- Sample B described below) for both line and continuum.
Shown are least squares linear fits to the gradient, $\lte\,(K) = a_1
+ a_2\, \rgal\,({\rm kpc}) $, for the entire sample and, separately, for
nebulae located inside and outside the solar orbit. The gradient is
flatter for \hii\ regions in the inner Milky Way.

Seven nebulae in Figure~\ref{fig:tegrad} are flagged because they do
not follow the general \lte\ gradient.  Two nebulae, G49.582$-$0.38
(1,851 K, 6.5 kpc) and G5.956$-$1.265 (3,416 K, 7.8 kpc) have
temperatures much lower than the $7,585 \pm 1,262$ K average \lte\ for
the 6--8 kpc interval of \rgal.  Moreover, the G49.582$-$0.38
temperature is $\sim$\,3,600 K lower than the Wink et al. (1983)
value.  For these sources we had to use TP continuum measurements to
derive \lte.  This may have compromised the accuracy of our result.
G5.899$-$0.427 (11,128 K, 6.1 kpc) also lies far from the general
trend of the sample. Here, however, we do not have any reason to
suspect that its \lte\ derivation is less reliable than that of other
sources in the 6--8 kpc zone.

The four remaining anomalous nebulae are located at the extremes of
the \rgal\ distribution of our \hii\ region sample.  These nebulae
thus have a large influence on the temperature gradient fits.  The two
Galactic Center sources, G1.13$-$0.1 (7,135 K, 0.1 kpc) and Sgr~B2
(8,169 K, 0.4 kpc), may well share the anomalous chemical abundances
in this region (see \S\,\ref{sec:gc}).  The two outer Galaxy sources,
S209 (10,506 K, 16.9 kpc) and S209~N (12,565 K, 16.7 kpc), are part of
the same \hii\ region.  Here again we can identify no compelling
reason to exclude these nebulae.

We analyzed 6 subsets of our \hii\ region sample in order to assess
the effects of sample selection on the temperature 
gradient.\footnote{G345.40$+$1.41, S291 and S132 were excluded from all
samples.  We have no continuum information for G345.40$+$1.41.  The
other nebulae have large uncertainties associated with either their
line or continuum data.}
Sample A includes all 109 of our sources.  
Sample B includes only high QF sources (the 78 sources shown in 
Figure~\ref{fig:tegrad} which have QFs {\it C}\, or better).
Sample C removes TP continuum sources from sample B for a total
of 64 nebulae.
Sample D removes the 7 sources flagged in Figure~\ref{fig:tegrad} 
from Sample B for a total of 71 nebulae.
Sample E removes G49.582$-$0.38 and G5.956$-$1.265 from sample B 
for a total of 76 nebulae.
Sample F removes 3 sources from sample E (the two Galactic Center
sources and G5.899$-$0.427) for a total of 73 nebulae.  

We fit least squares linear temperature gradients to each sample.  The
fit results are summarized in Table~\ref{tab:tefit} which lists the
coefficients $a_1$ and $a_2$ and their the standard deviations $\sigma
(a_1)$ and $\sigma (a_2)$, together with the correlation coefficient,
$r$, the $\chi ^2$ of the fit, the number, $N$, of sources in the
sample, and the \rgal\ range of the sample, $\Delta\,\rgal$.  In all
cases the gradient is close to 300 \kkpc.  Even our smallest sample of
64 objects is astrophysically significant in the sense that it spans a
large range of Galactic radius.  
The fit to the sample E nebulae is shown as a solid line in
Figure~\ref{fig:tegrad}.  {\it Unless otherwise stated we use the
sample E gradient fit in all subsequent analysis described herein.}
This fit is $T_e = (5780 \pm 350) + (287 \pm 46) \rgal\,$ ($r = 0.59$,
$N=76$).

Figure~\ref{fig:tegrad} shows \lte\ error bars.  Many nebulae have
error bars which are smaller than the plotted symbols. The error bars
shown are the propagated statistical errors in the measured
quantities. Various systematic effects almost certainly lead to
uncertainties larger than the error bars shown. Because systematic
errors are inherently unquantifiable and certainly are not normally
distributed, any attempt to plot ``systematic error bars'' would be
misleading.

This is especially true for errors in the kinematically determined
\rgal.  They are entirely dominated by systematic effects including
Galactic scale streaming motions, differences in the choice of
rotation curve, etc. Estimates for the magnitude of the error of the
\rgal\ determination are typically $\sim$\,25\% (e.g. Kuchar \& Bania
1990).  
This value is the quadrature sum of the $\sim$\,15\% error 
due to the uncertainty in the rotation curve and of the $\sim$\,20\% 
error due to streaming/non-circular motions.    
Here we have, however, at least made the \rgal\ determinations
in a uniform, systematic way that distinguishes this effort from the
majority of \te\ gradient studies which normally draw their distances
from a heterogeneous mixture of techniques.

It is important to understand the nature of the scatter in
Figure~\ref{fig:tegrad}. Deharveng et al. (2000) argued that a
significant fraction of the scatter in earlier investigations of the
electron temperature gradient may result from observational errors and
sample inhomogeneity. They cite the case of the bright \hii\ region
S206: the electron temperature estimated by various authors ranges
from $ \leq 8,000$ K (Churchwell et al. 1978; Mezger et al. 1979) to
about 13,000 K (Litchen et al. 1979). In our study sample homogeneity
is not a problem and the statistcal errors are quite small compared to
the scatter in \lte.

\begin{figure}
\epsscale{0.75} \plotone{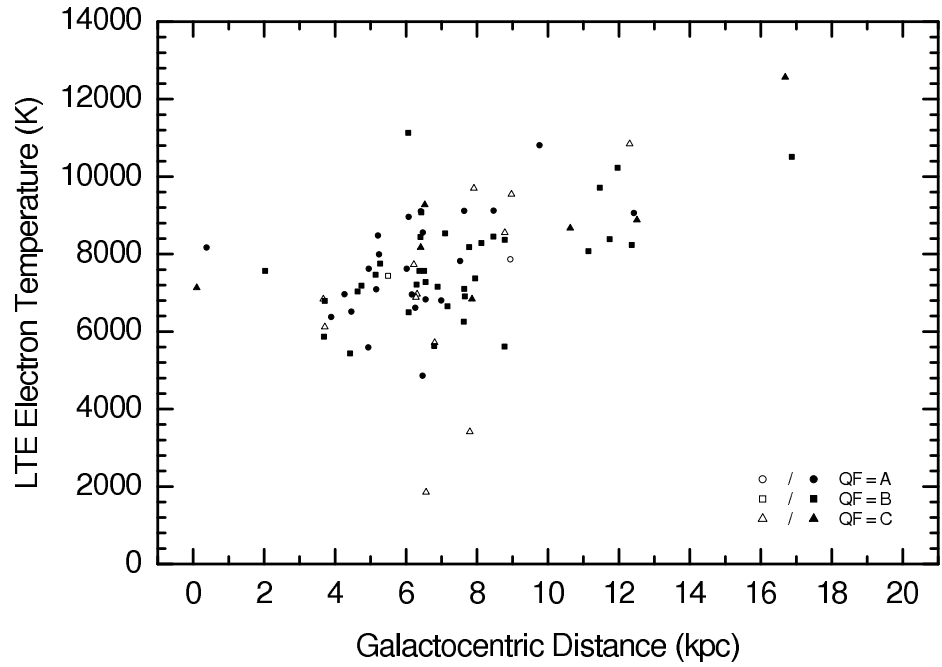} \figcaption[tergalqfcont]
{Repeat of Figure~\ref{fig:tegrad} showing the quality factor of 
the continuum data.  Filled symbols denote \lte\ temperatures derived
from switched power continuum measurements;  open symbol sources used 
total power measurements.
\label{fig:te-qf}}
\end{figure}

Could systematic error be responsible for the scatter?  We believe
that the main source of systematic error arises from determining the
baseline level of our continuum measurements. The continuum QF's givem
at least a qualitative estimate of this.  In Figure~\ref{fig:te-qf} we
plot the Figure~\ref{fig:tegrad} points without error bars and use
different symbols to identify the continuum QFs. The two lower
outliers are indeed QF = {\it C}, but for the bulk of the points the
differing QF's yield a comparable scatter. {\it From this we conclude
that the scatter in the points indicates a real spread in \lte\/.}
That is, the dispersion in \lte\ at any \rgal\ is indicative of 
intrinsic variations in \lte\ between nebulae.

\subsection{Electron Temperature Variations in Galactic Radius} 
\label{sec:radtegrad}

\begin{figure}
\epsscale{0.75} \plotone{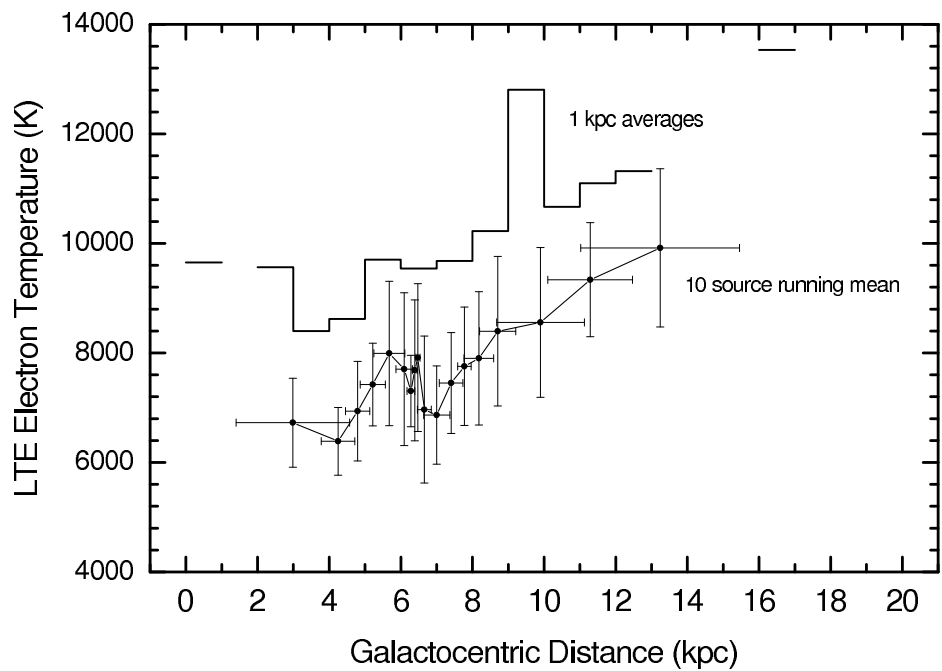} \figcaption[f5.lower.ps]
{``Smoothed'' electron temperatures plotted as a function of
Galactocentric distance, \rgal.  Symbols with error bars are a 10
point running mean of \lte\ and \rgal\ drawn at intervals of 4 points
along \rgal.  Horizontal line segments show the mean \lte\ in each 1
kpc wide \rgal\ interval. These are offset by +2,000 K for clarity.
Both smoothing algorithms suggest a slightly smaller gradient in the
inner Galactic disk.
\label{fig:sm-grad}}
\end{figure}

Given the size, uniformity, and precision of our sample, we can
investigate whether there is more complex behavior than a linear
gradient in the variation of \lte\ with \rgal. To reduce the scatter
to make trends more visible we have ``smoothed'' the data in two ways
as shown in Figure~\ref{fig:sm-grad}. The symbols with error bars are
a 10 point running mean of \lte\ and \rgal\ plotted at intervals of 4
points along Galactocentric distance.  The horizontal line segments
show the mean \lte\ in each 1 kpc interval, ranging from $\rgal =
0$--1 kpc to $\rgal = 16$--17 kpc. These are offset by +2,000 K for
clarity.  Both techniques smooth the data in slightly different ways;
both suggest a slightly smaller gradient in the inner area of the
Galactic disk.

We also made a least squares second order polynomial fit to the
electron temperature gradient for the sample E sources.  This fit was
very similar to the best linear fit, and perhaps even misleading
because of the effect of the radial outliers.  We therefore divided
sample E sources into nebulae inside and outside the solar orbit and
then fit two linear segments to these data subsets.  These fits are
shown in Figure~\ref{fig:tegrad}. The slope interior to \rgal\ is less
than that in the outer Galaxy: $153 \pm 85$ \kkpc\ compared to $404
\pm 130$ \kkpc. This result is strongly influenced by the outliers. If
we exclude G1.13$-$0.1, G5.899$-$0.427, Sgr~B2 and S209 from the
linear fits, we have a gradient of $268 \pm 96$ \kkpc\ for $\rgal <
\ro$ and $342 \pm 239$ \kkpc\ for $\rgal > \ro$.  We conclude that our
data do not justify anything more elaborate than a single linear
fit. The data do hint at more complex behavior, but a larger sample of
nebulae at both very small and very large \rgal\ are needed to explore
it.  
This is to be expected since our nebular sample was not chosen to
study the disk electron temperature gradient.  Indeed having more
nebulae at larger Galactocentric distances would significantly improve
the determination of the electron temperature gradient. We in fact
intend to make observations of more nebulae at larger \rgal\ in the
future because of this.  

\subsection{Electron Temperature Variations in Galactic Azimuth} 
\label{sec:azmtegrad}

\begin{figure}
\epsscale{0.75} \plotone{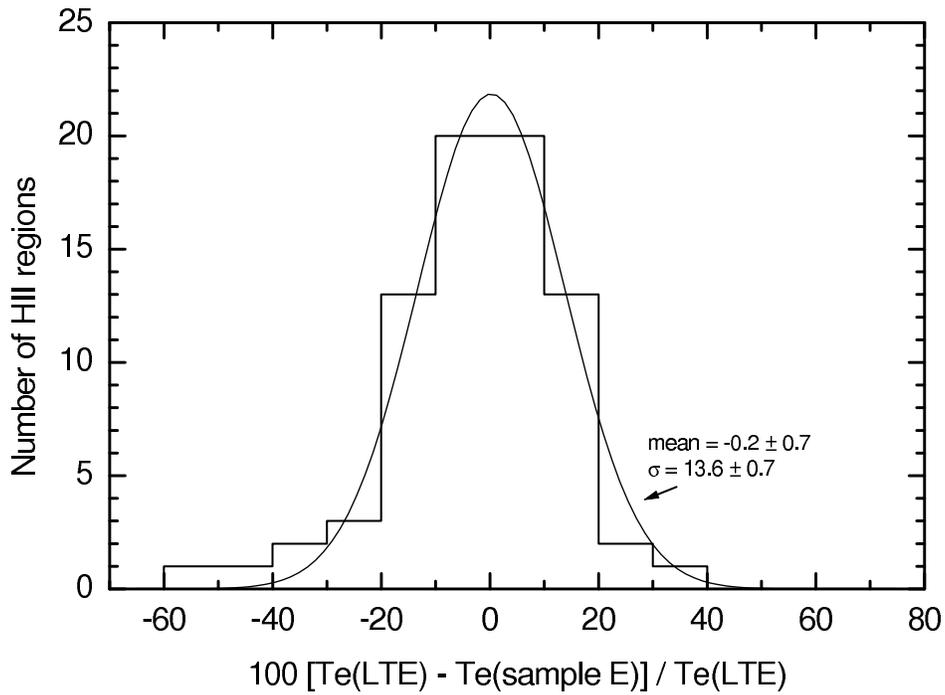} \figcaption[histog2]
{Histogram of the fractional deviation of the nebular electron
temperature from the best temperature gradient model fit to 
sample E. The deviations are well fitted by a Gaussian distribution 
whose dispersion (sigma) corresponds to an intrinsic 
electron temperature fluctuation of $\sim$\,1,100 K at any 
Galactic radius.
\label{fig:delta-te}}
\end{figure}

In \S~\ref{sec:magtegrad} we conclude that the scatter in \lte\ at a
given \rgal\ is not due to observational error. Figure~\ref{fig:delta-te} 
shows a histogram of the percentage deviation of the nebular \lte\
from the best \te\ gradient model fit to sample E.  The deviations are
Gaussian distributed with a $\Delta T_e/T_e$ dispersion of about 14\%.  
This implies that Galactic \hii\ regions have intrinsic \lte\ fluctuations 
of $\sim$\,1,100 K at any \rgal.
Shaver et al. (1983) also argued that most of their scatter in
$T_e$ is intrinsic. Their Figure~17 shows that a realistic range in
the effective temperature of the exciting stars (30,000--45,000 K) or
of the electron density can account for a spread of $T_e$ as large as
2,000 K; this is confirmed by photoionization models (Rubin 1985;
see \S\,\ref{sec:hiite}).

Maps of radio continuum emission (Altenhoff et al. 1978; Reich et al.
1990; F\"urst et al. 1990; Haynes et al. 1978), and our own continuum
observations show that many \hii\ regions are found spatially close to
each other. The gas within a given complex of nebulae could share the
same nucleosynthetic history perhaps including self-enrichment.  The
nucleosynthetic history might, however, vary significantly from
complex to complex.  We searched our \hii\ region sample for spatially
clumped clusters of nebulae with similar \lte\ values.  We found no
obvious signature of patchy nucleosynthesis.

Another possibility is that the assumption of axial symmetry in the
stellar production of heavy elements is invalid. If the radial
gradient were a function of Galactocentric azimuth, then the scatter
in a \lte\ vs \rgal\ Figure~\ref{fig:tegrad}--type plot would result
from lumping together nebulae from different azimuths into the same
\rgal\ bin.  We have therefore searched for azimuthal differences in
our nebular sample.

\begin{figure}
\epsscale{0.75} \plotone{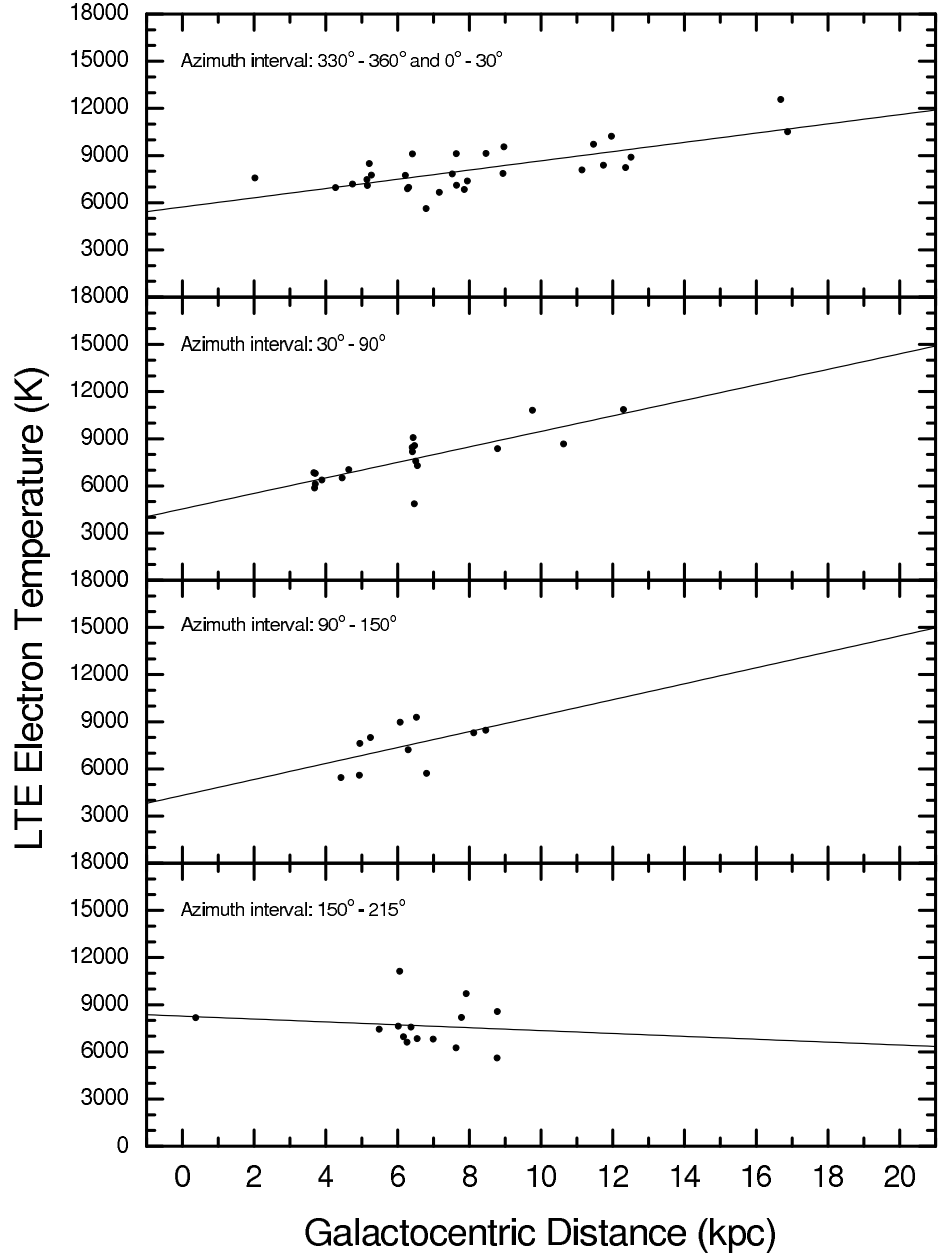}\figcaption[f6.azmgrad.eps]
{Nebular electron temperature plotted as a function of Galactocentric
radius for four distinct ranges of Galactocentric azimuth.  The analysis 
is compromised because our \hii\ region sample is not uniformly 
distributed in the Milky Way's disk (see Figure~\ref{fig:azm}).
\label{fig:azmgrad}}
\end{figure}

Figure~\ref{fig:azmgrad} shows the nebular electron temperature
plotted as a function of Galactocentric radius for four distinct
ranges of Galactocentric azimuth.  A difficulty in the analysis of the
azimuthal variation of the temperature gradient is that our sample is
not uniformly distributed in the disk.  Some azimuth intervals have
large concentrations of \hii\ regions (e.g. 350\arcdeg--20\arcdeg),
while others have no objects at all (see Figure \ref{fig:azm}).  We
divided our \hii\ region sample into four azimuth ranges chosen such
that each contains a comparable number of nebulae:
300\arcdeg--30\arcdeg, 30\arcdeg--90\arcdeg, 90\arcdeg--150\arcdeg,
and 150\arcdeg--215\arcdeg \ (G49.582$-$0.38 and G5.956$-$1.265 are not
included in this analysis).

Most of the intervals are relatively well populated between \rgal\
2--10 kpc. Because of this we fit the \lte\ gradient over a shorter
\rgal\ interval, roughly from 3--9 kpc.  Properties of the gradient
fits are given in Table~\ref{tab:azmfit} for each azimuth interval.
(Table~\ref{tab:azmfit} gives the same fit information as
Table~\ref{tab:tefit}.)  There is a variation in the \lte\ gradient
derived for the different azimuth ranges.  It is impossible, however,
for us to draw any significant astrophysical inferences from this
result because the azimuth ranges do span different angular zones,
cover different \rgal\ ranges, and contain different numbers of
nebulae.

In sum, we find no definitive evidence for clumpiness or azimuthal
variations in the distribution of nebular electron temperatures.  A
much larger \hii\ region sample that is more uniformly distributed in
the Milky Way disk is needed.

\subsection{Comparison With Other Milky Way Electron Gradient Studies} 
\label{sec:compgrad}

We searched the literature for previous determinations of the electron
temperature gradient in the Galactic disk.  Table~\ref{tab:gradcomp}
summarizes the results of these efforts.  Besides the reference for
and gradient found by each study, Table~\ref{tab:gradcomp} lists the
Galactocentric distance interval together with the assumed radius of
the solar orbit, $\Delta$\,\rgal\,(kpc) (${\rm R_0}$\,[kpc]), the
number, $N$, of objects, as well as the type of observation and
analysis method used to derive the gradient.  Most studies of
gradients in electron temperature in the Milky Way are based on
observations of radio recombination lines.  These RRL gradients vary
from 250--440 \kkpc.  Wink et al. (1983) suggest a gradient of 270
\kkpc\ based on observations of the 76\,$\alpha$ line for 84
\hii\ regions. 

Shaver et al. (1983) derived electron temperature and abundance
gradients which have often been used as the basis of many models for
the chemical evolution of our Galaxy (Tosi 1988; Giovagnoli \& Tosi
1995; Thon \& Meusinger 1998; among others). Because their gradient,
$433 \pm 40$ \kkpc, is one of the steepest values ever determined,
here we try to understand why this is so.

Different adopted distances are one possible source of disageeement.
The Shaver et al.  distances were derived from the measured radial
velocities using the Schmidt (1965) rotation curve with $\ro = 10$
kpc.  According to the authors their \rgal\ distances should be
accurate to within 1--2 kpc in most cases. There are a few objects in
common with our sample which have \rgal\ differences much larger than
that expected from the change in Galactic rotation parameters and
rotation curves.  For example, G1.13$-$0.1 is placed at $\rgal \approx
6.0$ kpc in Shaver et al. whereas we derive a distance of $\rgal
\approx 0.1$ kpc. Rudolph et al. (1997) give $\rgal \approx 0.91$ kpc
for this \hii\ region (Lis 1991, Simpson \& Rubin 1990). Distances for
G0.6$-$0.6 are also quite different. One way to compare our results to
Shaver et al.'s is to consider only the group of 22 \hii\ regions in
common. If we adopt their distances for these 22 sources our electron
temperature gradient steepens from $258 \pm 50$ \kkpc\ to $363 \pm 37$
\kkpc\ which is still far short of the $433 \pm 40$ \kkpc\ Shaver et
al. result.

The Shaver et al. source sample consists of 67 distinct Galactic \hii\
regions located in the range \rgal\ = 3.5--13.7 kpc. The RRL data for
44 \hii\ regions were their own 5 GHz (H\,109\,$\alpha$) observations.  
This was supplemented by 14.7 GHz (H\,76\,$\alpha$) observations
of 23 southern \hii\ regions by McGee \& Newton (1981). All these
observations were made with the Parkes 64 m radio telescope.

\begin{figure}
\epsscale{0.75} \plotone{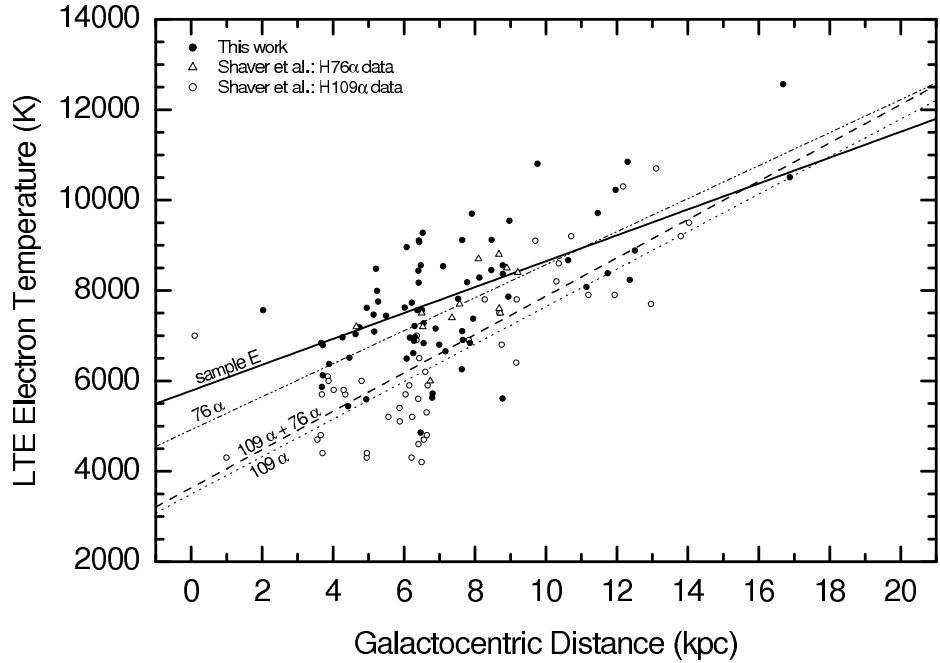}\figcaption[f7.s83.eps]
{Comparison of our \hii\ region electron temperatures with those from
Shaver et al. (1983).  The Shaver et al. nebulae were observed with
the H\,76\,$\alpha$\,(triangles) and H\,109\,$\alpha$\,(circles)
recombination lines.  Their sample has a large number of objects in
the interval \rgal\ = 2--6 kpc.  Their  H\,109\,$\alpha$\, \lte\ values 
in this zone are systematically lower than ours which may contribute 
to the very steep electron temperature gradient they derive.
\label{fig:s83}}
\end{figure}

We compare our \lte\ results with those of Shaver et al. in
Figure~\ref{fig:s83} where temperatures derived from the H$\,76\alpha$
and H$\,109\alpha$ lines are plotted with different symbols (triangles
and circles, respectively). To avoid the distance issue, we used their
measured velocities to derive new kinematic values for \rgal\ in
exactly the same way that we derived our own distances. We show only
those points for which we could recompute the Shaver et al. \rgal. Our
new distances (as well as most of the original Shaver, et
al. distances) were derived using Galactic rotation curves derived
from northern hemisphere data. North-South symmetries in Galactic
rotation have long been known (e.g., Burton 1988), so it might not be
appropriate to use our adopted rotation curve for distances to fourth
quadrant \hii\ regions. Fourth quadrant sources make up most of the
Shaver et al. sample for $\rgal > 6$ kpc.

Figure~\ref{fig:s83} shows that the electron temperatures obtained
using the H\,76$\alpha$ data are completely consistent with our
results. The discordance between our work and that of Shaver et
al. (1983) arises mostly from the low temperature points calculated
with the H$\,109\alpha$ lines located between $\rgal = 4$--6
kpc. Deharveng et al. (2000) also noted that the Shaver et al. (1983)
electron temperatures in the \rgal\ = 3--7 kpc zone are generally
lower than those derived by other RRL studies (Mezger et al. 1979;
Wink et al. 1983; Caswell \& Haynes 1987).

There are at least three factors which could steepen the Shaver et al.
(1983) gradient: a systematic difference between the \te\ derived from
the H\,76$\alpha$ and H$\,109\alpha$ lines; a change in distance
scales at roughly $\rgal > 6$ kpc; or a different abundance gradient
for Fourth Quadrant \hii\ regions.\footnote{We find an indication of
just this for our Fourth Quadrant nebular sample shown in
Figure~\ref{fig:azmgrad}.}

We also compare our results with those by Afflerbach et al. (1996),
who measured electron temperatures in ultracompact \hii\ regions,
using the H\,42\,$\alpha$, H\,66\,$\alpha$, H\,76\,$\alpha$ and
H\,93\,$\alpha$ RRLs. They found a Galactocentric gradient $\te\,[K] =
(320 \pm 64)\,\rgal\ + (5537 \pm 387)$ which is in good agreement with
our value despite the fact that ultracompact \hii\ regions are much
denser nebulae than our sources.

Optical studies of Galactic scale \hii\ region electron temperature
gradients are difficult because dust in the Galactic plane causes high
extinction in the visible.  Optical observations by Peimbert et
al. (1978) gave a large gradient of $\sim\,1100$ \kkpc. Their result
was probably influenced by the short \rgal\ interval spanned by their
sample. A more recent optical estimate by Deharveng et al. (2000)
yields a gradient consistent with that gotten by RRL methods.
Finally, radio continuum observations at low radio frequency by Omar
et al. (2002) are also consistent with the gradient estimated by
Deharveng et al. (2000) for the \rgal\ interval spanning 10--18 kpc
(albeit for a much smaller sample).

Maciel \& Fa\'undez-Abans (1985) investigated the radial electron
temperature gradient for a large sample of Peimbert (1978) Type II
planetary nebulae (PNe) using electron temperatures derived from
forbidden lines of [\oiii].  Because Type II PNe have approximately
circular orbits and thus do not appreciably change their
Galactocentric distances during their lifetimes, they are well suited
to abundance gradient studies.  The observed scatter in \te\ 
is larger than for \hii\ regions, probably due to: the large range of
effective temperatures of the central stars, winds, optical depth
effects, etc.  Nonetheless, Maciel \& Fa\'undez-Abans (1985) found a
correlation between the electron temperature and Galactocentric
distance for type II PNe with a gradient of the order of 600 \kkpc\
and an uncertainty of about 20\%.

Compared to these much older PNe, \hii\ regions are of zero age and
thus sample the physical state of the current interstellar medium.
Thus the flattening of the \te\ vs \rgal\ gradient seen when one
compares the PNe and \hii\ regions may be caused by Galactic scale
temporal chemical evolution.  This time flattening of the electron
temperature gradient should be accompanied by a corresponding time
flattening of the abundance gradient.  In fact, such a flattening
based on the comparison of data from several types of objects
(planetary nebulae, \hii\ regions, open clusters, cepheids and young
stars) is proposed by Maciel et al. (2003; 2005; 2006).  It is also
predicted by some recent inside-out formation scenario chemical
evolution models (Hou et al. 2000; Alib\'es et al. 2001).

\section{DISCUSSION}\label{sec:discuss}

\subsection{The Electron Temperature of Galactic H\thinspace II Regions}
\label{sec:hiite}

The electron temperature of an \hii\ region in thermal equilibrium is
set by the balance of competing heating and cooling mechanisms.  It is
therefore somewhat surprising that there is a Galactic \te\ gradient.
There are at least four physical properties that could effect \te: (1)
the effective temperature of the ionizing star, $T_{\rm eff}$, which
sets the hardness of the radiation field exciting (and heating) the
nebula; (2) the electron density---collisional de-excitation in the
high $n_{e}$ \hii\ region will inhibit cooling and increase \te;
(3) dust grains which effect the heating and cooling in complex ways;
and (4) heavy element abundance which increases cooling and decreases
\te\ (Garay \& Rodr\'{\i}guez 1983).  

Rubin (1985) explored how metallicity, gas density, and the stellar
effective temperature effect the average electron temperature in model
\hii\ regions.  These models predict changes in \te\ of
7,000\kel\ for a factor of 10 change in metal abundance, 2,900\kel\
for a change in density from 100 to \expo{5}\percc, and 1,300\kel\ for
a change in $T_{\rm eff}$ from 33,000 to 45,000\kel\ (B0 to O5
spectral type).  Dust grains are known to play a significant role in
the heating and cooling of \hii\ regions (e.g., Mathis 1986; Baldwin
et al. 1991; Shields \& Kennicutt 1995).  Photoelectric heating occurs
as electrons are ejected from dust grains while the gas is cooled by
the collisions of fast particles with grains.  The electron
temperature will decrease with distance from the star as the ionizing
radiation field is attenuated by dust grains.  But the electron
temperature will also increase as coolants are depleted onto dust
grains.  Taking these competing factors into account Oliveira \&
Maciel (1986) conclude that dust grains do not significantly contribute
to the observed \te\ gradient with a maximum variation of 500\kel.
Therefore metallicity is the most sensitive factor that sets the
nebular \te\ value and so \hii\ region metal abundance variations are
the best interpretation for the observed \te\ gradient.

The electron temperature need not be a constant, however, and the
observational methods used to determine \te\ are sensitive to
different regions of the nebula.  The two main methods used to
determine electron temperatures in \hii\ regions are (1) recombination
line-to-continuum ratios, such as hydrogen RRL and continuum emission;
and (2) forbidden line ratios, such as the ratio
[\oiii]$\lambda$\,4363 / ($\lambda$\,4959 + $\lambda$\,5007).  The
recombination line method is weighted towards the lower temperature
regions with a weak dependence on \te, while the forbidden line method
is weighted towards the higher temperature regions with a strong
dependence on \te\ (e.g., Peimbert 1967).  Therefore if temperature
structure exists in \hii\ regions these methods can produce different
electron temperatures.  Observations of [\oiii] and oxygen
recombination lines in Orion show such temperature fluctuations
(Esteban et al. 1998).

Comparison of hydrogen RRL and [\oiii] electron temperatures is more
complicated since the emission lines arise from different species.
The intensities of the RRLs of H relative to the underlying continuum,
for example, should give an estimate of the electron temperature in
the whole ionized H region (H$^+$ zone), while observations of the
[\oiii] optical emission should give an estimate of the electron
temperature in the O$^{++}$ zone.  Photoionization models indicate
that the electron temperature should be higher in the outer regions
since the ionizing radiation becomes harder with distance from the
exciting star and the very efficient coolants like O$^{++}$ are
located close to the exciting star (Stasi\'nska 1980; Garnett
1992). This suggests that $T_e({\rm O}^{++}) \leq T_e({\rm H}^+) \leq
T_e({\rm O}^+)$ (Stasi\'nska 1990; Stasi\'nska \& Shaerer 1997;
Deharveng et al. 2000).  The temperature measured via the RRL,
however, is not strictly $T_e({\rm H}^+)$ and that measured with the
optical [\oiii] forbidden lines is not strictly $T_e({\rm
O}^{++})$. Because the recombination lines are emitted preferentially
in low-temperature regions, $T_e ({\rm radio}) \leq T_e({\rm H}^+)$,
and because the optical forbidden lines are enhanced in high
temperature regions, $T_e$([\oiii])$ \ge T_e({\rm O}^{++}$).  Despite
this complexity the electron temperatures determined by these two
methods towards the same \hii\ region produce values of \te\ that are
similar to within the uncertainties (Deharveng et al. 2000).

Wink et al. (1983) find no correlation between \te\ and either the
Lyman continuum flux, $N_{\rm L}$, used to probe different stellar
effective temperatures, or the electron density, $n_{e}$.
Theoretically, higher values of $N_{\rm L}$ and $n_{e}$ should
increase \te\ (Rubin 1985).  But there are \hii\ regions with high
$N_{\rm L}$ and low $n_{e}$ (e.g., the Rosette nebula) that will
increase the scatter in any such analysis.  Shaver et al. (1983) found
a correlation with \te\ by dividing the \hii\ regions into two groups
of either high or low values of {\it both} $N_{\rm L}$ and $n_{e}$.
But the correlation seems weak since many of the points at smaller
\rgal\ where the correlation is best are nebulae that appear to have
systematically lower electron temperatures (i.e. the H\,109\,$\alpha$
survey sources).  The ultracompact \hii\ region survey of Afflerbach
et al. (1996) probes nebula of much higher electron density than other
surveys.  Because these observations were made with an interferometer
any diffuse gas within their primary beam will be spatially filtered.
Their derived electron temperatures are higher by $\sim$\,1,000\kel\
compared with other surveys of classical \hii\ regions.

We explored the effects of $N_{\rm L}$ and $n_{e}$ on electron
temperature in our sample and find no correlation.  We used an
approach similar to Shaver et al. (1983) wherein we selected sources
either below or above a threshold value of the excitation and electron
density ($\log({\rm N_{\rm L}}) = 49.5\,{\rm s^{-1}}$ and $n_{e} = 150$\percc).
RRL and continuum surveys of classical \hii\ regions with single-dish
telescopes probe \hii\ region complexes that typically contain more
than one ionizing star with different spectral types.  Such regions
consist of both low and high density gas (e.g., Balser et al. 1999).
But does this account for the observed dispersion of the \te\ gradient
as has been suggested (e.g., Shaver et al 1983)?

Another possibility is that the dispersion is caused by variations in
metallicity.  After all, stellar abundances at a given \rgal\ show a
dispersion in abundance. Interpretation of the stellar abundances is
complicated because samples typically contain stars of varying age
which may well have been formed at a different \rgal\ than their
current location (Edvardsson et al. 1993). Correcting for birth
location and age Edvardsson et al. (1993) find a scatter of 0.5 dex in
the solar neighborhood.  Friel et al. (2002) find a scatter of about
half that in disk star cluster abundances. A scatter of 0.3 dex in
heavy element abundance will produce a scatter of 2,000 K in \te,
entirely consistent with our observed scatter.

For \hii\ regions, Mehringer et al. (1993) suggest differences in
metallicity as the best candidate for the $\sim 2,000$\kel\ change in
\te\ between components 1 and 2 in the Sgr\,B complex.  Since our
survey spans a large region of the Galaxy, both radially and
azimuthally, we were able to explore such effects by comparing \te\ at
a given \rgal\ over a large range of azimuth.  The results were
inconclusive.  Sensitive RRL observations in critial areas of the
Galaxy could reveal such variations and place important constraints on
Galactic chemical evolution models.

\subsection{The Galactic Center}\label{sec:gc}

Our nebular sample F fit to the electron temperature gradient excludes
the Galactic Center (GC) sources Sgr\,B2 and G1.13$-$0.1 and includes
the outer Galaxy nebulae S209 and S209\,N.  This sample produces the
steepest \lte\ gradient, thus providing an extreme constraint on
Galactic chemical evolution models.

Early RRL observations of \hii\ regions in the GC produced electron
temperatures that were higher than those expected from an
extrapolation of the electron temperature gradient in the Galactic
disk (Mezger et al. 1979; Downes et al. 1980; Wink et al. 1983).
Since the chemical evolution of the GC may be different than the disk,
it is not surprising that electron temperatures in the GC do not
follow the gradient in the disk.  Higher spatial resolution
observations reveal a wide range of electron temperatures ($\sim
4,000-10,000\,$K) in Galactic Center \hii\ regions (Roelfsema et
al. 1987; Gaume \& Claussen 1990; Mehringer et al. 1992,1993; De Pree
et al. 1995, 1996; Lang et al. 1997, 2001).  The large range of
electron temperatures may be due to the complexity of the GC.  Real
variations in metallicity, electron density, and excitation will
effect $T_{e}$.  Some of the higher electron temperatures are
overestimated when the radio continuum is contaminated with
non-thermal emission.  Nevertheless, in some objects very high
electron temperatures ($T_{e} \sim 20,000\,$K) appear to be real
where the gas may not be in equilibrium (Mehringer et al. 1995).
Since the zero level for our radio continuum temperature was
determined away from the thermal object but not far from the Galactic
disk, any smooth non-thermal emission should have been removed.

\subsection{The Galactic O/H Abundance Gradient}\label{sec:OtoH}

\begin{figure}
\epsscale{0.75} \plotone{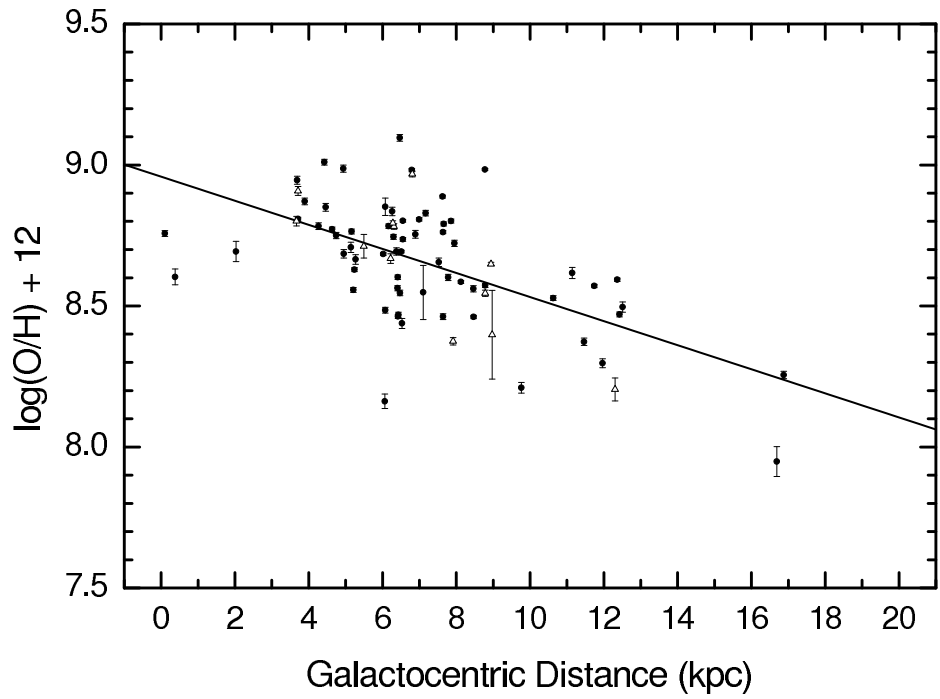} \figcaption[ohgrad]
{The Galactic O/H abundance gradient derived from our nebular electron
temperatures using the relation between O/H and \te\ of Shaver et al. (1983).  
Symbols have the same meaning as in Figure~\ref{fig:tegrad}.  The solid line
is the Eqn.~\ref{eq:ohgradient} least squares fit.
\label{fig:ohgrad}}
\end{figure}

The electron temperature may be converted to oxygen abundance using a
correlation between metal abundance and temperature. Because of the
broad interest in abundance gradients, we provide here one such
conversion of our \lte\ gradient to an O/H gradient, deferring a more
extended discussion to a future paper. Using the relation between O/H
and $T_e$ of Shaver et al. (1983) yields the result shown in
Figure~\ref{fig:ohgrad}. The O/H gradient is about $-$0.04 dex
kpc$^{-1}$ in agreement with results from Afflerbach et al. (1996),
Deharveng et al. (2000), Pilyugin (2003) and Esteban et al. (2005). 
The exact fit to the 76 sample E nebulae in Figure~\ref{fig:ohgrad} 
is:  
\begin{equation}\label{eq:ohgradient}
{\rm log\,(O/H) + 12 = (8.958 \pm 0.052)~+~(-0.043 \pm 0.007)\,\rgal\,(kpc)}
\end{equation}
with a correlation coefficient of $r=-0.59$.  If our observed \lte\
dispersion of $\sim$\,1,100 K is caused by real abundance
fluctuations, then this would correspond to an [O/H] abundance
dispersion of $\sim$\,0.16 dex.


\section{SUMMARY} \label{sec:summ}

We used extremely sensitive radio recombination line observations of a
large sample of Galactic \hii\ regions to study the nebular electron
temperature gradient in the Milky Way disk.  LTE electron temperatures
were derived from the observed H\,91\,$\alpha$ RRL line-to-continuum
intensity ratios.  Departures from LTE were found to be small given
the properties of our \hii\ regions and the fact that at 8.6 GHz the
effect of pressure broadening is not large. Our 109 source sample
consists mostly of classical \hii\ regions of low density and with
angular sizes larger than the $3\farcm20$ telescope beam.

Our temperature gradient analysis has the following virtues:
(1) We analyze a large sample of \hii\ regions spanning the entire
Galactic disk from \rgal\ = 0--17 kpc.  
(2) We use high sensitivity radio recombination line measurements 
obtained with very long integration times for our derivation of the 
electron temperature.
(3) All nebular Galactocentric distances are calculated kinematically 
in a self-consistent way.
(4) All of the data were observed with the same telescope, the NRAO 
140~Foot, identically calibrated, and analyzed in a uniform, self-consistent
way.

Our best estimate of the electron temperature gradient is derived from
a sample of 76 sources for which we have the highest quality data
(sample E of Table~\ref{tab:tefit}). We conclude that the present
gradient in electron temperature in the Galactic disk has a minimum at
the Galactic Center and rises at a rate of 287 $\pm$ 46 \kkpc.  This
value is consistent with determinations in the optical by Deharveng et
al. (2000) and in the radio by Wink et al. (1983), Afflerbach et
al. 1996 and Azc\'arate et al. 1985. Our gradient is about a factor of
two less steep than the gradient proposed by Shaver et al. (1983).

We find little if any variation of the electron temperature gradient
with Galactocentric distance.  There is some variation of the
temperature gradient calculated for different regions of
Galactocentric azimuth.  Unfortunately our nebular sample is not
homogeneously distributed in the Galactic plane which complicates
this analysis so no firm conclusions can be drawn.

The scatter we find in the \hii\ region electron temperatures at a
given Galactocentric radius is not due to observational error, but
rather to intrinsic fluctuations in these temperatures that are almost
certainly due to fluctuations in the nebular heavy element abundances.
A comparison of the \hii\ region gradient with much steeper gradient
found for planetary nebulae suggests that the electron temperature
gradient evolves with time, becoming flatter as a consequence of the
chemical evolution of the Milky Way's disk.

\acknowledgments

We thank the staff of NRAO Green Bank for their help, support and
friendship. The \he3\ research has been sporadically supported by the
National Science Foundation.  The most recent grants were AST 00-98047
to TMB and AST-0098449 to RTR.  We thank Butler Burton and Eileen
Friel for discussions that have improved this effort.  
The perceptive comments made by our anonymous referee helped us
to improve this paper.  
CQ is grateful to the Department of Astronomy at the University of
Virginia, for their hospitality. Her work was partly supported by the
Levinson Fund of the Peninsula Community Foundation, Funda\c c\~ao de
Amparo \`a Pesquisa do Estado de S\~ao Paulo (FAPESP) and Conselho
Nacional de Desenvolvimento Cient\'{\i}fico e Tecnol\'ogico (CNPq/MCT).

\clearpage



\begin{deluxetable}{llrrrrrrrrrrrrr}
\tabletypesize{\scriptsize}
\rotate
\tablecaption{Physical Properties of Galactic \hii\ Regions. \label{tab:physparam}}
\tablewidth{0pt}
\tablehead{
\colhead{Source} & \colhead{Name} & \colhead{\rgal} & \colhead{\dsun} &
\colhead{$\Theta_{\rm diam}$} & \colhead{D} & 
\colhead{${n(^4{\rm He}^+) / n({\rm H}^+)}$} & \colhead{$S_{\nu}$} & 
\colhead{${T_C}^B$} & \colhead{$n_e$} &
\colhead{$T_e^{\ast}$\tablenotemark{a}} &
\colhead{Mode\tablenotemark{b}} & \colhead{QFs\tablenotemark{c}} & \colhead{Survey} \\
\colhead{} & \colhead{} & \colhead{(kpc)} & \colhead{(kpc)} & 
\colhead{(\arcmin)} & \colhead{(pc)} & \colhead{} & \colhead{(Jy)} &
\colhead{(K)} & \colhead{(cm$^{-3}$)} & 
\colhead{(K)} &  \colhead{} & \colhead{}
& \colhead{} &\colhead{} 
}
\startdata
G0.605$+$0.325                      &   \nodata          &               11.5   &             20.0   &              5.7   &             33.2   &            0.065 $\pm$            0.025   &              0.8   &              0.2   &             13.2   &             6600 $\pm$              320   &     TP               &   B                E &   $^3$He           \\  
G0.665$-$0.035                      &   Sgr~B2           &                0.4   &              8.9   &              4.8   &             12.3   &            0.038 $\pm$            0.014   &             40.3   &             17.3   &            194.7   &             8170 $\pm$              180   &     SP               &   A                C &   $^3$He           \\  
G1.13$-$0.1                         &   \nodata          &                0.1   &              8.5   &              7.6   &             18.7   &            0.065 $\pm$            0.008   &             10.6   &              1.8   &             50.0   &             7130 $\pm$               70   &     SP               &   C                B &   $^3$He           \\  
G2.90$+$0.0                         &   \nodata          &               10.7   &             19.2   &              7.9   &             44.2   &            0.052 $\pm$            0.014   &              1.3   &              0.2   &             10.1   &             4840 $\pm$              140   &     TP               &   C                D &   $^3$He           \\  
G3.270$-$0.101                      &   \nodata          &                5.5   &             14.0   &              7.4   &             30.0   &            0.077 $\pm$            0.011   &              4.0   &              0.7   &             24.9   &             7440 $\pm$              280   &     TP               &   B                C &   $^3$He           \\
\enddata
\tablenotetext{a}{The electron temperatures were rounded of to the nearest 10\,K.} 
\tablenotetext{b}{Mode refers to the continuum observing mode, where TP is for total power and SP is for switched power observations.} 
\tablenotetext{c}{QFs corresponds to the continuum and line quality factors, respectively.}
\tablenotetext{d}{M16, M17, NGC~6334, Rosette and S209 were observed at different positions in the $^3$He survey.}
\tablenotetext{e}{Observed in both \cii\ and $^3$He surveys.}
\end{deluxetable}

\clearpage

\begin{deluxetable}{lrrrll}
\tabletypesize{\scriptsize}
\tablecaption{Average Differences in Electron Temperature Determinations\tablenotemark{a}. \label{tab:tecomp2}}
\tablewidth{0pt}
\tablehead{
\colhead{Reference} & \colhead{N} &
\colhead{${\rm 100\,<(us - ref)/us>}$} & 
\colhead{${\rm 100\,<abs(us - ref)/us>}$}
& \colhead{Observation} & \colhead{Method}}
\startdata                                                          
PTR78                 &   2    &    3.6    &   12.9 & optical & [\oiii]\,$\lambda$4363,5007     \\
WBW79                 &   6    &    3.6    &    4.9 & RRL     & $T_L({\rm H}\,66\,\alpha)/T_C$  \\
TMP80                 &   7    &    2.9    &    4.7 & RRL     & $T_L({\rm H}\,66\,\alpha)/T_C$  \\
TMP80                 &   5    &    2.7    &   11.0 & RRL     & $T_L({\rm H}\,76\,\alpha)/T_C$  \\
GR83                  &   3    &    9.3    &   11.1 & RRL     & $T_L({\rm H}\,125\,\alpha)/T_C$ \\
WWB83                 &  30    &   10.9    &   14.4 & RRL     & $T_L({\rm H}\,76\,\alpha)/T_C$  \\
SMNDP83               &  17    &   13.3    &   14.2 & RRL     & $T_L({\rm H}\,109\,\alpha)/T_C$ \\
ACA96                 &   4    &    4.5    &   15.8 & RRL     & $T_L({\rm H}\,66\,\alpha,76\,\alpha,93\,\alpha)/T_C$ \\
\enddata
\tablenotetext{a}{QF = A, B and C, without the outlier G49.582$-$0.38.}
\tablerefs{
ACA96 - Afflerbach et al. (1996);
BBRW99 - Balser et al. (1999);
DPCC00 - Deharveng et al. (2000); 
EGPPRR05 - Esteban et al. (2005); 
GR83 - Garay \& Rodr\'{\i}guez (1983);
OCAR02 - Omar et al. (2002); 
PTR78 - Peimbert et al. (1978);
SMNDP83 - Shaver et al. (1983);
TMP80 - Thum et al. (1980);
WBW79 - Wilson et al. (1979);
WWB83 - Wink et al. (1983).
}
\end{deluxetable}

\clearpage

\begin{deluxetable}{lrrrrrr}
\tabletypesize{\scriptsize}
\tablecaption{\lte\ Gradient Fits: $T_e^* = a_1 + a_2 \rgal $  \label{tab:tefit}}
\tablewidth{0pt}
\tablehead{
\colhead{Sample} & \colhead{$a_1$(K)} & \colhead{$a_2$(\kkpc)} & \colhead{$r$} & \colhead{$\chi ^2$} & \colhead{$N$} & \colhead{$\Delta \rgal$(kpc)}
}
\startdata
A\tablenotemark{a} & 5840 $\pm$ 410 & 281 $\pm$ 49 & 0.49 & 1690 & 109 & 0.0--20.0 \\ 
B\tablenotemark{b} & 5660 $\pm$ 430 & 286 $\pm$ 56 & 0.50 & 1430 &  78 & 0.0--17.0 \\
C\tablenotemark{c} & 5960 $\pm$ 380 & 261 $\pm$ 50 & 0.56 & 1199 &  64 & 0.0--17.0 \\
D\tablenotemark{d} & 5480 $\pm$ 410 & 312 $\pm$ 55 & 0.56 & 1053 &  71 & 2.0--13.0 \\
E\tablenotemark{e} & 5780 $\pm$ 350 & 287 $\pm$ 46 & 0.59 & 1173 &  76 & 0.0--17.0 \\
F\tablenotemark{f} & 5300 $\pm$ 350 & 340 $\pm$ 45 & 0.67 & 1059 &  73 & 2.0--17.0 \\
\cutinhead{$\rgal < \ro$}
A\tablenotemark{a} & 6110 $\pm$  690 & 222 $\pm$  110 & 0.23 & 1675 & 75 &  0.0--8.5 \\
B\tablenotemark{b} & 6750 $\pm$  660 &  89 $\pm$  106 & 0.11 & 1430 & 62 &  0.0--8.5 \\
\enddata
\tablenotetext{a}{All QFs: 106 sources plus 6 additional pointings
within 5 of those sources. S291
and S132 were not included due to their large errors; we
have no continuum for G345.40+1.41.}
\tablenotetext{b}{QF = A, B and C.}
\tablenotetext{c}{Sample B excluding sources calculated with only TP continuum.}
\tablenotetext{d}{Sample B excluding outlying values of
${T_e}^{\ast}$ flagged in Figure~\ref{fig:tegrad}.}
\tablenotetext{e}{Sample B excluding the outliers G49.582$-$0.38 and G5.956$-$1.265.}
\tablenotetext{f}{Sample E, but excluding Sgr~B2, G1.13$-$0.1 and G5.899$-$0.427}
\end{deluxetable}

\clearpage

\begin{deluxetable}{lrrrrrr}
\tabletypesize{\scriptsize}
\tablecaption{\lte\ Gradient Fits: $T_e^* = a_1 + a_2 \rgal $ for Different Azimuth Intervals\tablenotemark{a}.\label{tab:azmfit}}
\tablewidth{0pt}
\tablehead{
\colhead{Azimuth Range} & \colhead{$a_1$(K)} & \colhead{$a_2$(\kkpc)} & \colhead{$r$} & \colhead{$\chi ^2$} & \colhead{$N$} & \colhead{$\Delta R$ (kpc)}
}
\startdata
\sidehead{330\arcdeg--30\arcdeg}
\nodata & 5730 $\pm$  480 & 294 $\pm$  53 & 0.73 & 997 & 29 & 2.0--16.9 \\
\nodata & 6070 $\pm$  1070 & 233 $\pm$  157 & 0.33 & 970 & 20 & 3.0--9.0 \\
\sidehead{30\arcdeg--90\arcdeg}
\nodata & 4540 $\pm$  650 & 493 $\pm$  95 & 0.79 & 1010 & 18 & 3.7--12.3 \\
\nodata & 4830 $\pm$  950 & 432 $\pm$  168 & 0.58 & 990 & 15 & 3.7--9.0 \\
\sidehead{90\arcdeg--150\arcdeg}
\nodata & 4320 $\pm$  2050 & 507 $\pm$  324 & 0.48 & 1320 & 10 & 4.4--8.5 \\	
\nodata & 4320 $\pm$  2050 & 507 $\pm$  324 & 0.48 & 1320 & 10 & 4.4--8.5 \\
\sidehead{150\arcdeg--215\arcdeg}
\nodata & 8270 $\pm$  1360 & $-$92 $\pm$  200 & $-$0.13 & 1481 & 14 & 0.4--8.8 \\
\nodata & 8500 $\pm$  2910 & $-$120 $\pm$  410 & $-$0.09 & 1546 & 13 & 3.0--8.8 \\
\enddata
\tablenotetext{a}{QF = A, B and C, without the outliers G49.582$-$0.38 and G5.956$-$1.265.}
\end{deluxetable}

\clearpage

\begin{deluxetable}{lrrrrr}
\tabletypesize{\scriptsize}
\tablecaption{Comparison of Electron Temperature Gradient Determinations \label{tab:gradcomp}}
\tablewidth{0pt}
\tablehead{
\colhead{Reference} & \colhead{$\Delta R$} & \colhead{$d T_e / d R$} & \colhead{$N$} & \colhead{Observation} & \colhead{Method} \\
}
\startdata
PTR78     &  8.4--13.9 (10.0)  &  11000             &        5         &      optical  &   [\oiii]\,$\lambda$4363,5007              \\         
LRC79     &  5.0--13.0 (10.0)  &  390$\pm$70        &       20         &      RRL      &   $T_L({\rm H}\,86\,\alpha)/T_C$                 \\        
LRC79     &  5.0--13.0 (10.0)  &  250$\pm$150       &       20         &      RRL      &   $T_L({\rm H}\,108\,\beta)/T_C$                 \\        
WPZ79     &  5.0--12.5 (10.0)  &  400$\pm$100       &       18         &      RRL      &   $T_L({\rm H}\,65\,\alpha)/T_C$                              \\              
DWBW80    &  4.0--09.0 (10.0)  &  340$\pm$90        &       115\tablenotemark{a}       &      RRL      &   $T_L({\rm H}\,110\,\alpha)/T_C$                             \\            
MN81      &  5.0--11.0 (10.0)  &  420               &       23         &      RRL      &   $T_L({\rm H}\,76\,\alpha)/T_C$                             \\             
GR83      &  3.6--12.6 (10.0)  &  440$\pm$50        &       23         &      RRL      &   $T_L({\rm H}\,125\,\alpha)/T_C$                             \\                        
SMNDP83   &  3.5--13.7 (10.0)  &  433$\pm$40        &       67         &      RRL      &   $T_L({\rm H}\,109\,\alpha,76\,\alpha)/T_C$                \\
\enddata
\tablenotetext{a}{Objects averaged in 6 groups corresponding to increments of 1 kpc in Galactocentric radius.}
\tablenotetext{b}{$T_e$([\oiii]) in 6 \hii\ regions plus mean temperatures obtained, in 1 kpc bins, from the following RRLs studies: Mezger et al. (1979), Shaver et al. (1983), Wink et al. (1983) and Caswell \& Haynes (1987).}
\tablerefs{
ACA96 - Afflerbach et al. 1996;
ACC85 - Azc\'arate et al. 1985;
DPCC00 - Deharveng et al. 2000;
DWBW80 - Downes et al. 1980; 
GR83 - Garay \& Rodr\'{\i}guez 1983;
LRC79 - Lichten et al. 1979;
MN81 -  McGee \& Newton 1981;
PTR78 - Peimbert et al. 1978;
SMNDP83 - Shaver et al. 1983;
WPZ79 - Wilson et al. 1979;
WWB83 - Wink et al. 1983.
}
\end{deluxetable}



\clearpage

\clearpage

\clearpage

\clearpage

\clearpage

\clearpage

\clearpage

\clearpage

\clearpage

\end{document}